\documentclass[12pt]{iopart}
\usepackage{amssymb}
\usepackage{txfonts}
\usepackage{graphicx}
\usepackage{subfigure}
\usepackage{dcolumn}
\usepackage{bm}
\usepackage{sidecap}

\begin{document}

\title[Source monitoring for continuous-variable quantum key distribution]{Source monitoring for continuous-variable quantum key distribution}

\author{Jian Yang, Xiang Peng$^{*}$ and Hong Guo$\dag$}

\address{CREAM Group, State Key Laboratory of Advanced Optical
Communication Systems and Networks (Peking University) and Institute
of Quantum Electronics, School of Electronics Engineering and
Computer Science, Peking University, Beijing 100871, PR China}

\ead{$^{*}$xiangpeng@pku.edu.cn}
\ead{$\dag$hongguo@pku.edu.cn}

\begin{abstract}
The noise in optical source needs to be characterized for the security of continuous-variable quantum key distribution (CVQKD).
Two feasible schemes, based on either active optical switch or passive beamsplitter are proposed
to monitor the variance of source noise, through which, Eve's knowledge can be properly estimated. We derive the security bounds for both
schemes against collective attacks in the asymptotic case, and find that the passive scheme performs better.
\end{abstract}

\maketitle
\section{Introduction}
Continuous-variable quantum key distribution (CVQKD) encodes information into the quadratures of
optical fields and extracts it with homodyne detection, which has
higher efficiency and repetition rate than that of the single photon
detector \cite{Scarani_RMP_09}. CVQKD, especially the GG02 protocol
\cite{Grosshans_PRL_02}, is hopeful to realize high speed
key generation between two parties, Alice and Bob.

Besides experimental demonstrations \cite{Grosshans_Nature_03, Lodewick_Ex_PRA_07},
the theoretical security of CVQKD has been established against collective attacks
\cite{Grosshans_PRL_05, Navascues_PRL_05}, which has been shown optimal in the
asymptotical limit \cite{Renner_PRL_102}. The practical security of CVQKD has also
been noticed in the recent years, and it has been shown that the source noise in
state preparation may be undermine the secure key rate \cite{Filip_PRA_08}. In GG02,
the coherent states should be displaced in phase space following Gaussian
modulation with variance $V$. However, due to the imperfections in laser source and
modulators, the actual variance is changed to $V+\chi_{s}$, where $\chi_{s}$
is the variance of source noise.

An method to describe the trusted source noise is the beamsplitter model
\cite{Filip_PRA_08, Filip_PRA_10}. This model has a good
approximation for source noise, especially when the transmittance of beamsplitter
$T_{A}$ approaches $1$, which means that the loss in signal mode is
negligible. However, this method has the difficulty of parameter
estimation to the ancilla mode of the beamsplitter, without the information of
which, the covariance matrix of the system are not able to determine. In this case,
the optimality of Gaussian attack \cite{Patron_PRL_06,Navascues_PRL_06} should be
reconsidered \cite{YujieShen__PRA_11}, and we have to assume that the channel is linear
to calculate the secure key rate.
To solve this problem, we proposed an improved source noise model
with general unitary transformation \cite{YujieShen__PRA_11}. Without extra
assumption on quantum channel and ancilla state, we are able to derive a tight
security bound for reverse reconciliation, as long as the variance of source
noise $\chi_{s}$ can be properly estimated. The optimality of Gaussian attack is kept
within this model.

The remaining problem is to estimate the variance of source noise
properly.  Without such a source monitor, Alice and Bob can not discriminate
source noise from channel excess noise, which is supposed to be controlled by
the eavesdropper (Eve) \cite{Yong_JPB_09}. In practice,
source noise is trusted and is not controlled by Eve. So, such
\textit{untrusted source noise model} just overestimates Eve'power and
leads to an untight security bound. A compromised method is to
measure the quadratures of Alice's actual
output states each time before starting experiment. However, this work is time consuming, and
in QKD running time, the variance of source noise may fluctuate slowly and deviate
from preliminary result. In this paper, we propose two real-time
schemes, that the active switch scheme and the passive beamsplitter
scheme to monitor the variance of source noise, with the help of which,
we derive the security bounds asymptotically for both of them against
collective attacks, and  discuss their potential applications when
finite size effect is taken into account.


\section{Source monitoring in CVQKD}
In this section, we introduce two real-time schemes to monitor the variance
of source noise for the GG02 protocol, based on our general model. Both 
schemes are implemented in the so-called
prepare and measurement scheme (P\&M scheme) \cite{Garcia_PHD_2009}, while for the
ease of theoretical research, here we analyze their security in
the entanglement-based scheme (E-B scheme). The covariance matrix, used to simplify the
calculation, is defined by
\cite{Garcia_PHD_2009}
\begin{equation}
\gamma_{ij}=\textrm{Tr}[\rho\{(\hat{r}_{i}-d_{i}), (\hat{r}_{j}-d_{j})\}],
\end{equation}
where operator $\hat{r}_{2i-1}=\hat{x}_{i}$, $\hat{r}_{2i}=\hat{p}_{i}$, mean value
$d_{i}=\langle\hat{r}_{i}\rangle=\textrm{Tr}[\rho\hat{r}_{i}]$, $\rho$ is
the density matrix, and $\{\}$ denotes the anticommutator.

In E-B scheme, Alice prepares EPR pairs, measuring the quadratures of one mode
with two balanced homodyne detectors, and then send the other mode to Bob. It is easy
to verify that the covariance matrix of an EPR pair is
\begin{equation}
\gamma_{AB_{0}}=\left(
                  \begin{array}{cc}
                    V\mathbb{I} & \sqrt{V^{2}-1}\sigma_{z} \\
                    \sqrt{V^{2}-1}\sigma_{z} & V\mathbb{I} \\
                  \end{array}
                \right),
\end{equation}
where $V=V_{A}+1$ is the variance of the EPR modes, and $V_{A}$
corresponds to Alice's modulation variance in the P\&M scheme.
However, due to the effect of source noise, the actual covariance
matrix is changed to
\begin{equation}
\gamma_{AB_{0}}=\left(
                  \begin{array}{cc}
                    V\mathbb{I} & \sqrt{V^{2}-1}\sigma_{z} \\
                    \sqrt{V^{2}-1}\sigma_{z} & (V+\chi_{s})\mathbb{I} \\
                  \end{array}
                \right),
\end{equation}
where $\chi_{s}$ is the variance of source noise. As mentioned in
\cite{YujieShen__PRA_11}, we assume this noise is introduced by a neutral
party, Fred, who purifies $\rho_{AB}$ and introduces the source noise
with arbitrary unitary transformation. In this section, we show how
to monitor $\chi_{s}$ with our active and passive schemes, and derive
the security bounds in the infinite key limit.

\subsection{Active switch scheme}
\begin{figure}[t]
\begin{center}
\includegraphics[height=1.5in]{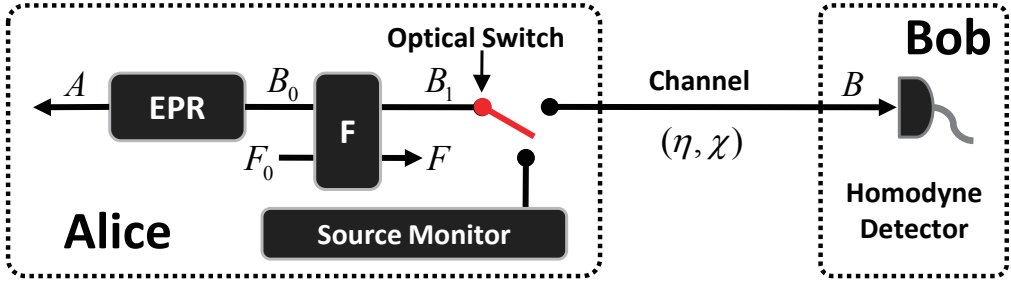}
\end{center}
\caption{(color online). Entanglement-based model for the optical switch scheme. Alice measures one mode of EPR pairs and projects the other mode to coherent states, and then sends it to Bob. F represents the neutral party, Fred, who introduces the source noise. Using a high-speed optical switch driven by TRNG, we can measure part of the signal sent to B and estimate the variance of source noise $\Delta V$ in the infinite key limit. }\label{pic2}
\end{figure}

A method of source monitoring is to use an active optical
switch, controlled by a true random number generator (TRNG),
combined with a homodyne detection. The
entanglement-based version \cite{Grosshans_QIC_2003} of this
scheme is illustrated in
Fig. \ref{pic2}, where we randomly select parts of signal pulses,
measure their quadratures and estimate their variance. In the
infinite key limit, the pulses used for source monitor should have
the same statistical identities with that sent to Bob. Comparing
the estimated value with the theoretical one, we are able to derive
the variance of source noise, and the security bound can be
calculated by \cite{YujieShen__PRA_11}
\begin{equation}
K_{OS}=(1-r)\times[\beta\times I(a:b)-S(E:b)],
\end{equation}
where $r$ is the sampling ratio of source monitoring,
$I(a:b)$ is the classical mutual information between Alice and
Bob, $S(E:b)$ is the quantum mutual information between Eve and Bob,
and $\beta$
is the reconciliation efficiency. After channel transmission, the
whole system can be described by covariance matrix $\gamma_{FAB}$
\begin{equation}
\gamma_{FAB}=
\left(
  \begin{array}{cccc}
    F_{11} & F_{12} & F_{13} & F'_{14} \\
    F_{12}^{T} & F_{22} & F_{23} & F'_{24} \\
    F_{13}^{T} & F_{23}^{T} & V\mathbb{I} & \sqrt{\eta (V^{2}-1)}\sigma_{z} \\
    (F'_{14})^{T} & (F'_{24})^{T} & \sqrt{\eta (V^{2}-1)}\sigma_{z} & \eta (V+\chi_{s}+\chi)\mathbb{I} \\
  \end{array}
\right),
\end{equation}
where $\chi_{s}$ is the variance of source noise, $2\times 2$ matrix $F_{ij}$
is related to Fred's two-mode state, $\eta$ is the transmittance and
$\chi=(1-\eta)/\eta+\epsilon$ is the channel noise and $\epsilon$ is the channel
excess noise. In practice, the covariance matrix can be estimated with experimental
data with source monitor and parameter estimation. Here, for the ease of calculation,
we assume that parameters $\eta$ and $\epsilon$ have known values.

Given $\gamma_{FAB}$, the classical mutual information $I(a:b)$ can
be directly derived, while $S(E:b)$ can not,
since the ancilla state $F$ is unknown in our general model.
Fortunately, we can substitute $\gamma_{FAB}$
with another state $\gamma'_{FAB}$ when calculating $S(E:b)$
\cite{YujieShen__PRA_11}, where
\begin{equation}\label{CM'}
\gamma '_{FAB}=
\left(
  \begin{array}{cccc}
    \mathbb{I} & 0 & 0 & 0 \\
    0 & \mathbb{I} & 0 & 0 \\
    0 & 0 & (V+\chi_{s})\mathbb{I} & \sqrt{\eta [(V+\chi_{s})^{2}-1]}\sigma_{z} \\
    0 & 0 & \sqrt{\eta [(V+\chi_{s})^{2}-1]}\sigma_{z} & \eta (V+\chi_{s}+\chi)\mathbb{I} \\
  \end{array}
\right),
\end{equation}
and we have shown that such substitution provides a tight bound
for reverse reconciliation.

Here, we have assumed that the pulses generated in Alice is i.i.d., and
the true random number plays an important role in this scheme,
without which, the sampled pulses may have different statistical characters
from signal pulses sent to Bob. The asymptotic performance of this scheme
will be analyzed in sec. III.

\subsection{Passive beam splitter scheme}
\begin{figure}[b]
\begin{center}
\includegraphics[height=1.5in]{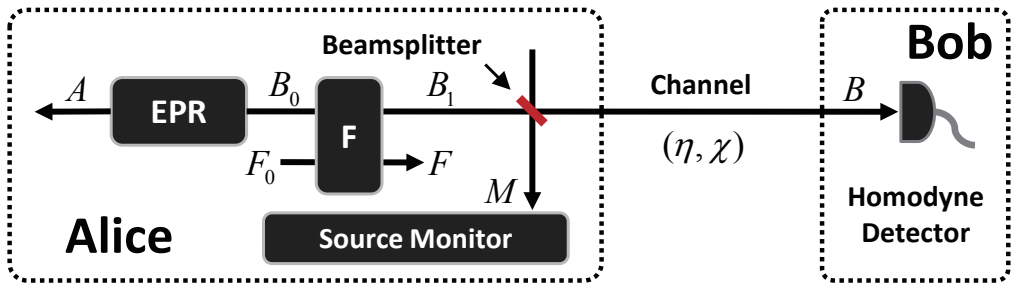}
\end{center}
\caption{(color online). Entanglement-based model for the beamsplitter scheme. The optical switch in Fig.\ref{pic2} is replaced by a beamsplitter. Alice and Bob are able to estimate the source noise by measuring mode $M$ with the homodyne detection. }\label{BSScheme}
\end{figure}

Though the active switch scheme is intuitive in theoretical research,
it is very not convenient in the experimental realization, since the high
speed optical switch and an extra TRNG are needed. Also,
it lowers the secure key rate with $(1-r)$ due
to the sampling ratio. Inspired by \cite{Peng_OL_2008}, we propose a
passive beam splitter scheme to simplify the implementation. As illustrated in
Fig. \ref{BSScheme}, a beamsplitter is used to separate mode
$B_{1}$ into two parts. One mode, $M$, is monitored by Alice, and the other,
$B_{2}$, is sent to Bob.

The security bound of passive beam splitter scheme can be calculated in a
similar way that we substitute the whole state $\rho_{FAB_{1}M_{0}}$ with
$\rho'_{FAB_{1}M_{0}}$. The covariance of its subsystem, $\rho_{AB_{1}M_{0}}$, can be written as
\begin{equation}
\gamma'_{AB_{1}M_{0}}=
\left(
  \begin{array}{ccc}
    (V+\chi_{s})\mathbb{I} & \sqrt{(V+\chi_{s})^{2}-1}\sigma_{z} & 0 \\
    \sqrt{(V+\chi_{s})^{2}-1}\sigma_{z} & (V+\chi_{s})\mathbb{I} & 0 \\
    0 & 0 & \mathbb{I} \\
  \end{array}
\right),
\end{equation}
where mode $M_{0}$ is initially in the vacuum state. The covariance matrix
after beam splitter is
\begin{equation}
\gamma'_{AB_{2}M}=(\mathbb{I}^{A}\otimes S_{\rm BS}^{BM})^{T}\gamma_{AB_{1}M_{0}}(\mathbb{I}^{A}\otimes S_{\rm BS}^{BM}),
\end{equation}
where
$$
\mathbb{I}^{A}\otimes S_{\rm BS}^{BM}=
\left(
  \begin{array}{ccc}
    \mathbb{I} & 0 & 0 \\
    0 & \sqrt{T}\mathbb{I} & \sqrt{1-T}\mathbb{I} \\
    0 & -\sqrt{1-T}\mathbb{I} & \sqrt{T}\mathbb{I} \\
  \end{array}
\right).
$$
Then, mode $B_{2}$ is sent to Bob through quantum channel,
characterized by $(\eta, \chi)$. The calculation of $S(E:b)$
in this scheme is a little more complex, since an extra mode $M$ is
introduced by the beamsplitter. We omit the detail of calculation here,
which can be derived from \cite{Garcia_PHD_2009}. The
performance of this scheme is discussed in the next section.


\section{Simulation and Discussion}
\begin{figure}[t]
\begin{center}
\includegraphics[height=3in]{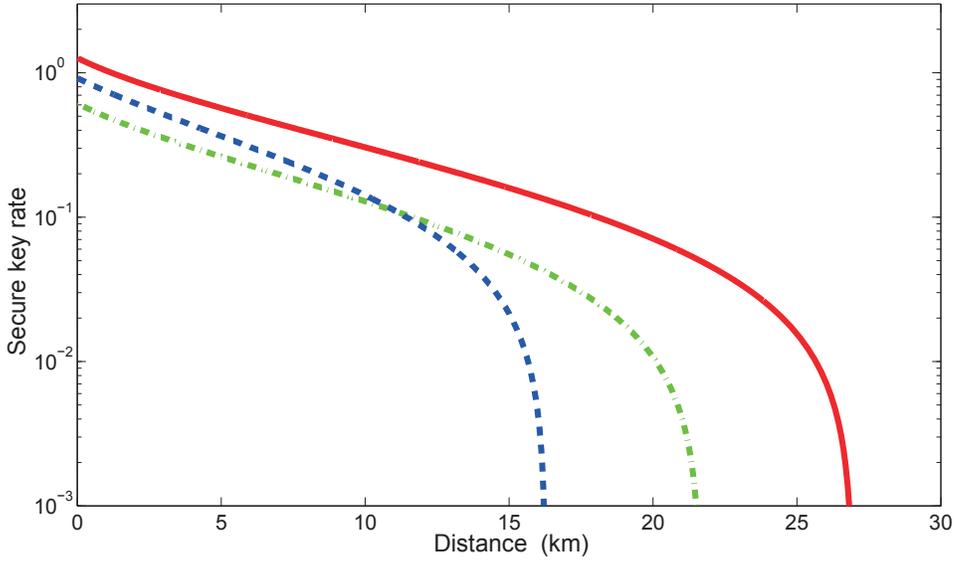}
\end{center}
\caption{(color online).  A comparison among the secure key rate
of untrusted noise scheme, optical switch scheme and beam splitter
scheme for the GG02 protocol, which are in dash line, dot-dash line and solid line,
respectively. Typical values are used for each parameter. The
modulation variance is $V=40$, the source noise is $\chi_{s}=0.1$,
the channel excess noise is $\epsilon=0.1$, and the reconciliation
efficiency is $\beta=0.8$. The sample ration in the optical switch
scheme is $r=0.5$, and the transmittance in the beam splitter scheme
is $T=0.5$.}\label{Comparision}
\end{figure}

In this section, we analyze the performance of both schemes with
numerical simulation.
As mentioned above, the simulation is restricted to the asymptotic
limit. The case of finite size will be
discussed later. To show the performance of source monitor schemes,
we illustrate the secure key rate
in Fig. \ref{Comparision}, in which the
\textit{untrusted noise scheme} is included for comparison. For the
ease of discussion, the imperfections in practical detectors are not
included in our simulation, the effect of which have been studied
previously \cite{Lodewick_Ex_PRA_07}.

As shown in Fig. \ref{Comparision}, secure key rate of each scheme
is limited within $30 {\rm km}$, where large excess noise
$\epsilon\sim 0.1$ is used. Under state-of-the-art technology, the
excess noise can be controlled less than a few percent of the shot
noise. So, our simulation is just a conservative estimation on the
secure key rate. The \textit{untrusted source noise scheme} has the
shortest secure distance, because it ascribes the source
noise into channel noise, which is supposed to be induced by the
eavesdropper. In fact, source noise is neutral and can be
controlled neither by Alice and Bob, nor by Eve.
So, this scheme just overestimates Eve's power by supposing
she can acquire extra information from source noise, which lower
the secure key rate of this scheme.

Both the active and passive schemes have longer secure distance
than the untrusted noise scheme, since they are based on the general
source noise model, which does not ascribe source noise into Eve's
knowledge. The active switch scheme has lower secure key rate in
the short distance area. This is mainly because that the random
sampling process intercepts parts of the signal pulses to estimate
the variance of source noise, which reduces the repetition rate with
ratio $r$. Nevertheless, it does not overestimate Eve's power
\cite{YujieShen__PRA_11}. As a result, the secure key distance is
improved.

\begin{figure}[t]
\begin{center}
\includegraphics[height=3.5in]{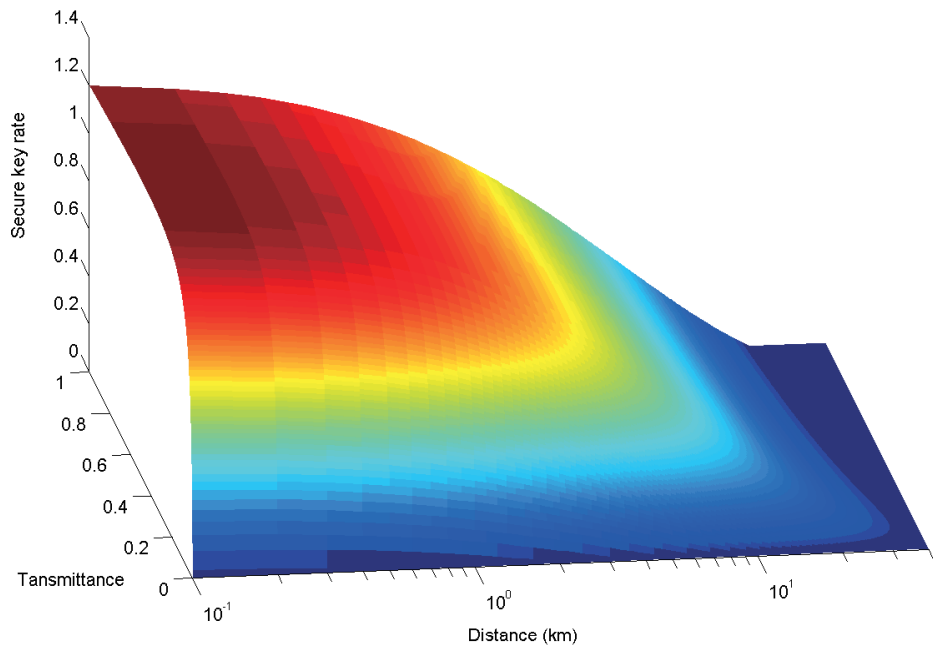}
\end{center}
\caption{(color online). Secure key rate as a function of distance $d$ and transmittance $T$, in which T varies from $0.01$ to $0.99$.
The colored parts illustrates the area with
positive secure key rate, the empty parts illustrates to insecure area, and the abscissa values of boundary points corresponds to the
secure distance.}\label{3D}
\end{figure}

Both the secure key rate and secure
distance of beam splitter scheme are superior than that of other
schemes, when the transmittance is set to be $0.5$, equal
to the sampling rate $r$ in optical switch scheme, where no extra
vacuum noise is introduced. This phenomena is quite similar to
the "noise beat noise" scheme \cite{Garcia_PHD_2009}, which
improves the secure key
rate by introduce an extra noise into Bob's side. Though such noise
lowers the mutual information between Alice and Bob, it also
makes Eve more difficult to estimate Bob's measurement result.
With the help of simulation, we find a similar phenomenon in the
beam splitter scheme, the vacuum noise reduces mutual information
$S(b:E)$ more rapidly than its effect on $I(a:b)$. A preliminary
explanation is that the sampled pulse in optical switch scheme is
just used to estimate the noise  variance, while in beam splitter
model it increases Eve's uncertainty on Bob's information. Combined
with advantages in experimental realization, beam splitter scheme
should be a superior choice.

To optimize the performance of passive beam splitter scheme,  we
illustrate the secure key rate in Fig. \ref{3D} for
different beam splitter transmittance $T_{A}$. The maximal secure
distance about $34{\rm km}$ is achieved when
$T\sim 0.1$, about $10$ km longer than that when $T\sim 0.5$.
Combined with the discussion above, this result
can be understood as a balance between the effects of the noise
 on $I(a:b)$ and $S(b:E)$, induced by beamsplitter. When $T_{A}$
is too small, $I(a:b)$ also decreases rapidly, which limits the
secure distance.

\section{Finite size effect}
The performance of source monitor schemes above is analyzed in
asymptotical limit. In practice, the real-time monitor will
concern the finite-size effect, since the variance of source noise
may change slowly.  A thorough research in finite size effect is beyond
the scope of this paper, because the security of CVQKD in finite size is still
under development, that the optimality of Gaussian attack and collective attack
has not been shown in the finite size case. Nevertheless, we are able to give a
rough estimation on the effect of block size, for a given distance. Taking the
active optical switch scheme for example, with a similar
method in \cite{Leverrier_PRA_2010}, the maximum-likelihood estimator
$\hat{\sigma_{s}}^{2}$ is given by
\begin{equation}
\hat{\sigma_{s}}^{2}=\left(\frac{1}{m}\sum_{i=1}^{m}y_{i}^{2}-V\right),
\end{equation}
where $(m\hat{\sigma_{s}}^{2}/\sigma_{s}^{2})\sim\chi^{2}(m-1)$,
$y_{i}$ is the measurement result of source monitor,
and $\sigma_{s}^{2}=\chi_{s}$ is the expected value of the variance
of source noise. For large $m$, the
$\chi^{2}$ distribution converges to a normal distribution. So, we have
\begin{equation}
\sigma_{\rm min}^{2}\approx \hat{\sigma}_{s}^{2}-z_{\epsilon_{\rm SM}}\frac{\hat{\sigma}_{s}^{2}\sqrt{2}}{\sqrt{m}}
\end{equation}
where $z_{sm}$ is such that $1-{\rm erf}(z_{sm}/\sqrt{2})=\epsilon_{\rm SM}$, and
$\epsilon_{SM}$ is the failure probability. The reason why we choose
$\hat{\sigma}_{min}$ is that given the values of $\eta(V+\chi_{s}+\chi)$ and $\eta$,
estimated by Bob, the minimum of $\chi_{s}$ corresponds to the maximum of channel noise $\chi$,
which may be fully controlled by Eve. The extra variance $\Delta_{m}\chi_{s}$ due to the
finite size effect in source monitor is
\begin{equation}\label{eq:block size}
\Delta_{m}\chi_{s}\approx \frac{z_{\rm SM}\hat{\sigma}_{s}^{2}\sqrt{2}}{\sqrt{m}}.
\end{equation}
For $\epsilon_{SM}\sim 10^{-10}$, we have $z_{\epsilon_{SM}}\approx 6.5$. As analyzed
in \cite{Leverrier_PRA_2010}, if the distance between Alice and bob is $50$ km
($T\sim 10^{-1}$), The block length should be at least $10^{8}$, which
corresponds to  $\Delta_{m} \chi_{s}\sim 10^{-6}$ induced by
the finite size effect in source monitor. Compared with the channel excess noise of $10^{-2}$,
the effect of finite size in source monitor is very slight. Due to the high repetition rate in
CVQKD, Alice and Bob are able to accumulate such a block within several minutes, during
which the source noise may change slightly.

\section{Concluding Remarks}

In conclusion, we propose two schemes, the active optical switch
scheme and the passive beamsplitter scheme, to monitor the variance
of source noise $\chi_{s}$. Combined with previous general noise model,
we derive tight security bounds for both schemes with reverse
reconciliation in the asymptotic limit. Both schemes can be implemented
under current technology, and the simulation result shows an better
performance of our schemes, compared with the untrusted source noise model.
Further improvement in secure distance can be achieved, when
the transmittance $T_{A}$ is optimized.

In practise, the source noise varies slowly. To realize
real-time monitoring, the finite size effect should be taken into account,
that the block size should not be so large, that the source noise has changed
significantly within this block, and the block size should not be too small,
that we can not estimate the source noise accurately.  The security proof
of CVQKD with finite block size has not been established completely, since the
optimality of collective attack and Gaussian attack has not been shown in finite
size. Nevertheless, we derive the effective source noise induced by the finite
block size, and find its effect is not significant in our scheme. So, our
schemes may be helpful to realize real-time source monitor in the future.

\section*{Acknowledgments}

This work is supported by the Key Project of National Natural
Science Foundation of China (Grant No. 60837004), National Hi-Tech
Research and Development (863) Program. The authors thank Yujie Shen,
Bingjie Xu and Junhui Li for fruitful discussion.

\end{document}